\documentstyle[12pt]{article}
\begin{document}

\title{\Large\bf Gauging the nonlinear sigma-model through a
non-Abelian algebra}

\author{J. Barcelos-Neto\thanks{\noindent e-mail:
barcelos@if.ufrj.br} ~and W. Oliveira\thanks{\noindent e-mail:
wilson@fisica.ufjf.br -- Permanent address: UFJF -  Departamento de
F\'{\i}sica - Juiz de Fora, MG 36036-330}\\
Instituto de F\'{\i}sica\\ 
Universidade Federal do Rio de Janeiro\\
RJ 21945-970 - Caixa Postal 68528 - Brasil\\
\date{}}

\maketitle
\abstract
We use an extension of the method due to Batalin, Fradkin, Fradkina,
and Tyutin (BFFT) for transforming the nonlinear $\sigma$ model in a
non-Abelian gauge theory. We deal with both supersymmetric and
nonsupersymmetric cases. The bosonic case was already considered in
literature but just gauged with an Abelian algebra. We show that the
supersymmetric version is only compatible with a non-Abelian gauge
theory. The usual BFFT method for this case leads to a nonlocal
algebra.

\vfill
\vspace{1cm}
\newpage

\section{Introduction}

\bigskip
Batalin, Fradkin, Fradkina, and Tyutin (BFFT) \cite{BFF,BT} developed
an elegant formalism of transforming systems with second-class in
first-class ones, i.e. in gauge theories \cite{Dirac}. This is
achieved with the aid of auxiliary fields that extend the phase space
in a convenient way to transform the second-class into first-class
constraints. The original theory is matched when the so-called
unitary gauge is chosen.

\medskip
In the way that the BFFT method was originally formulated, the so
obtained first-class constraints are imposed to form an Abelian
algebra. This is naturally the case of systems with linear
second-class constraints. Recently, Banerjee {\it et
al}~\cite{Banerjee1}, studying the non-Abelian Proca model, have
adapted the BFFT method in order that first-class constraints can
form a non-Abelian algebra
\footnote {For systems with initial first and second-class
constraints, the former has also to be modified in order to keep the
same initial algebra, which can be abelian or non-Abelian
\cite{Kim}.}. 
From these examples, it might appear that the original formulation of
the BFFT method is only addressed to theories with linear
second-class constraints while the Banerjee {\it et al.} extension,
to nonlinear ones. In fact, gauge theories obtained from systems with
linear constraints are always Abelian
\footnote {We mention that the BFFT method when applied to these
theories is equivalent to express the dynamical quantities by means
of linear shifted fields \cite{Ricardo}.}.
However, concerning the nonlinear ones we mention that the same
non-Abelian Proca model, the non-linear sigma model, and the
CP$^{N-1}$ have been recently studied in the context of the original
BFFT formalism \cite{Banerjee2,Banerjee3,Banerjee4}. In spite of
this, it is important to emphasize that the possibility pointed out
by Banerjee {\it et al.} to obtain non-Abelian gauge theories leads
to a richer structure comparing with the usual BFFT case.

\medskip
The purpose of the present paper is to convert the nonlinear $\sigma$
model into a non-Abelian gauge theory (the CP$^{N-1}$ could be done
in a similar way). This is an elucidating example in order to compare
the two faces of the method. We shall see for example that the
supersymmetric version is only consistently transformed in a gauge
theory by means of a non-Abelian algebra, while its Abelian
counterpart is nonlocal. Another interesting point is that the
obtained theory is a kind of Liouville one, for both supersymmetric
and nonsupersymmetric cases.

\medskip
Our paper is organized as follows. In Sec. 2 we make a brief review
of the BFFT method and its extension in order to emphasize and
clarify some of its particularities that shall be used in the
forthcoming sections. We also take this opportunity to include some
recent improvements of the method. In Sec. 3 we consider the BFFT
with the first-class constraints forming a non-Abelian algebra for
the nonlinear $\sigma$ model. Its supersymmetric formulation is
considered in Sec. 4. Concluding observations are given in Sec. 5.

\vspace{1cm}
\section{Brief review of the BFFT formalism and its recent
improvements} 
\renewcommand{\theequation}{2.\arabic{equation}}
\setcounter{equation}{0}

\bigskip
Let us consider a system described by a Hamiltonian $H_0$ in a
phase-space $(q^i,p^i)$ with $i=1,\dots,N$. Here we suppose that the
coordinates are bosonic (extensions to include fermionic degrees of
freedom and to the continuous case can be done in a straightforward
way). It is also supposed that there just exist second-class
constraints. Denoting them by $T_a$, with $a=1,\dots ,M<2N$, we have

\begin{equation}
\bigl\{T_a,\,T_b\bigr\}=\Delta_{ab}
\label{2.1}
\end{equation}

\bigskip\noindent
where $\det(\Delta_{ab})\not=0$. 

\medskip
As was said, the general purpose of the BFFT formalism is to convert
second-class constraints into first-class ones. This is achieved by
introducing canonical variables, one for each second-class constraint
(the connection between the number of second-class constraints and
the new variables in a one-to-one correlation is to keep the same
number of the physical degrees of freedom in the resulting extended
theory). We denote these auxiliary variables by $\eta^a$ and assume
that they have the following general structure

\begin{equation}
\bigl\{\eta^a,\,\eta^b\bigr\}=\omega^{ab}
\label{2.2}
\end{equation}

\bigskip\noindent
where $\omega^{ab}$ is a constant quantity with
$\det\,(\omega^{ab})\not=0$. The obtainment of $\omega^{ab}$ is
embodied in the calculation of the resulting first-class constraints
that we denote by $\tilde T_a$. Of course, these depend on the new
variables $\eta^a$, namely

\begin{equation}
\tilde T_a=\tilde T_a(q,p;\eta)
\label{2.3}
\end{equation}

\bigskip\noindent 
and it is considered to satisfy the boundary condition

\begin{equation}
\tilde T_a(q,p;0)=\tilde T_a(q,p)
\label{2.4}
\end{equation}

\bigskip\noindent
The characteristic of these new constraints in the BFFT method, as it
was originally formulated, is that they are assumed to be strongly
involutive, i.e.

\begin{equation}
\bigl\{\tilde T_a,\,\tilde T_b\bigr\}=0
\label{2.5}
\end{equation}

\bigskip\noindent
The solution of Eq.~(\ref{2.5}) can be achieved by considering
$\tilde T_a$ expanded as

\begin{equation}
\tilde T_a=\sum_{n=0}^\infty T_a^{(n)}
\label{2.6}
\end{equation}

\bigskip\noindent
where $T_a^{(n)}$ is a term of order $n$ in $\eta$. Compatibility
with the boundary condition~(\ref{2.4}) requires 

\begin{equation}
T_a^{(0)}=T_a
\label{2.7}
\end{equation}

\bigskip
The replacement of Eq.~(\ref{2.6}) into~(\ref{2.5}) leads to a set of
equations, one for each coefficient of $\eta^n$. We list some of them
below:

\begin{eqnarray}
&&\bigl\{T_a,T_b\bigr\}
+\bigl\{T_a^{(1)},T_b^{(1)}\bigr\}_{(\eta)}=0
\label{2.8}\\
&&\bigl\{T_a,T_b^{(1)}\bigr\}+\bigl\{T_a^{(1)},T_b\bigr\}
+\bigl\{T_a^{(1)},T_b^{(2)}\bigr\}_{(\eta)}
+\bigl\{T_a^{(2)},T_b^{(1)}\bigr\}_{(\eta)}=0
\label{2.9}\\
&&\bigl\{T_a,T_b^{(2)}\bigr\}
+\bigl\{T_a^{(1)},T_b^{(1)}\bigr\}_{(q,p)}
+\bigl\{T_a^{(2)},T_b\bigr\}
+\bigl\{T_a^{(1)},T_b^{(3)}\bigr\}_{(\eta)}
\nonumber\\
&&\phantom{\bigl\{T_a^{(0)},T_b^{(2)}\bigr\}_{(q,p)}}
+\bigl\{T_a^{(2)},T_b^{(2)}\bigr\}_{(\eta)}
+\bigl\{T_a^{(3)},T_b^{(1)}\bigr\}_{(\eta)}=0
\label{2.10}\\
&&\phantom{\bigl\{T_a^{(0)},T_b^{(2)}\bigr\}_{(q,p)}+}
\vdots
\nonumber
\end{eqnarray}

\bigskip\noindent 
The notation $\{,\}_{(q,p)}$ and $\{,\}_{(\eta)}$, represents the
parts of the Poisson bracket $\{,\}$ relative to the variables
$(q,p)$ and $(\eta)$, respectively.

\medskip
Equations above are used iteratively in the obtainment of the
corrections $T^{(n)}$ ($n\geq1$). Equation~(\ref{2.8}) shall give
$T^{(1)}$. With this result and Eq.~(\ref{2.9}), one calculates
$T^{(2)}$, and so on. Since $T^{(1)}$ is linear in $\eta$ we may
write

\begin{equation}
T_a^{(1)}=X_{ab}(q,p)\,\eta^b
\label{2.11}
\end{equation}

\bigskip\noindent 
Introducing this expression into Eq.~(\ref{2.8}) and using
Eqs.~(\ref{2.1}) and (\ref{2.2}), we get

\begin{equation}
\Delta_{ab}+X_{ac}\,\omega^{cd}\,X_{bd}=0
\label{2.12}
\end{equation}

\bigskip\noindent 
We notice that this equation does not give $X_{ab}$ univocally,
because it also contains the still unknown $\omega^{ab}$. What we
usually do is to choose $\omega^{ab}$ in such a way that the new
variables are unconstrained. It might be opportune to mention that
sometimes it is not possible to make a choice like that \cite{Barc2},
In this case, the new variables are constrained. In consequence, the
consistency of the method requires an introduction of other new
variables in order to transform these constraints also into
first-class. This may lead to an endless process. However, it is
important to emphasize that $\omega^{ab}$ can be fixed anyway.

\medskip
However, even one fixes $\omega^{ab}$ it is still not possible to
obtain a univocally solution for $X_{ab}$. Let us check this point.
Since we are only considering  bosonic coordinates~\footnote{The
problem also exists for the fermionic sector.}, 
$\Delta_{ab}$ and $\omega^{ab}$ are antisymmetric quantities. So,
expression (\ref{2.12}) compactly represents $M(M-1)/2$ independent
equations.  On the other hand, there is no prior symmetry involving
$X_{ab}$ and they consequently represent a set of $M^2$ independent
quantities.

\medskip
In the case where $X_{ab}$ does not depend on ($q,p$), it is easily
seen that $T_a+\tilde T_a^{(1)}$ is already strongly involutive for
any choice we make and we succeed in obtaining $\tilde T_a$. If this
is not so, the usual procedure is to introduce $T_a^{(1)}$ into Eq.
(\ref{2.9}) to calculate $T_a^{(2)}$ and so on. At this point resides
a problem that has been the origin of some developments of the
method, including the adoption of a non-Abelian constraint algebra.
This occurs because we do not know {\it a priori} what is the best
choice we can make to go from one step to another. Sometimes it is
possible to figure out a convenient choice for $X_{ab}$ in order to
obtain a first-class (Abelian) constraint algebra in the first stage
of the process \cite{Banerjee3,Banerjee4}. It is opportune to mention
that in the work of reference \cite{Banerjee1}, the use of a
non-Abelian algebra was in fact a way of avoiding to consider higher
order of the iterative method. More recently, the method has been
used (in its Abelian version) beyond the first correction
\cite{Banerjee2} and we mention that sometimes there are problems in
doing this \cite{Barc1}.

\medskip
Another point of the usual BFFT formalism is that any dynamic function
$A(q,p)$ (for instance, the Hamiltonian) has also to be properly
modified in order to be strongly involutive with the first-class
constraints $\tilde T_a$. Denoting the modified quantity by $\tilde
A(q,p;\eta)$, we then have

\begin{equation}
\bigl\{\tilde T_a,\,\tilde A\bigr\}=0
\label{2.13}
\end{equation}

\bigskip\noindent
In addition, $\tilde A$ has also to satisfy  the boundary condition

\begin{equation}
\tilde A(q,p;0)=A(q,p)
\label{2.14}
\end{equation}

\bigskip
The obtainment of $\tilde A$ is similar to what was done to get
$\tilde T_a$, that is to say, we consider an expansion like

\begin{equation}
\tilde A=\sum_{n=0}^\infty A^{(n)}
\label{2.15}
\end{equation}

\bigskip\noindent 
where $A^{(n)}$ is also a term of order $n$ in $\eta$'s.
Consequently, compatibility with Eq.~(\ref{2.14}) requires that

\begin{equation}
A^{(0)}=A
\label{2.16}
\end{equation}

\bigskip\noindent 
The combination of Eqs.~(\ref{2.6}), (\ref{2.7}), (\ref{2.13}),
(\ref{2.15}), and (\ref{2.16}) gives the equations

\begin{eqnarray}
&&\bigl\{T_a,A\bigr\}
+\bigl\{T_a^{(1)},A^{(1)}\bigr\}_{(\eta)}=0
\label{2.17}\\
&&\bigl\{T_a,A^{(1)}\bigr\}+\bigl\{T_a^{(1)},A\bigr\}
+\bigl\{T_a^{(1)},A^{(2)}\bigr\}_{(\eta)}
+\bigl\{T_a^{(2)},A^{(1)}\bigr\}_{(\eta)}=0
\label{2.18}\\
&&\bigl\{T_a,A^{(2)}\bigr\}
+\bigl\{T_a^{(1)},A^{(1)}\bigr\}_{(q,p)}
+\bigl\{T_a^{(2)},\bigr\}
+\bigl\{T_a^{(1)},A^{(3)}\bigr\}_{(\eta)}
\nonumber\\
&&\phantom{\bigl\{T_a^{(0)},A^{(2)}\bigr\}_{(q,p)}}
+\bigl\{T_a^{(2)},A^{(2)}\bigr\}_{(\eta)}
+\bigl\{T_a^{(3)},A^{(1)}\bigr\}_{(\eta)}=0
\label{2.19}\\
&&\phantom{\bigl\{T_a^{(0)},A^{(2)}\bigr\}_{(q,p)}+}
\vdots
\nonumber
\end{eqnarray}

\bigskip\noindent
which correspond to the coefficients of the powers 0, 1, 2, etc. of
the variable $\eta$ respectively. It is just a matter of algebraic
work to show that the general expression for $A^{(n)}$ reads

\begin{equation}
A^{(n+1)}=-{1\over n+1}\,\eta^a\,\omega_{ab}\,X^{bc}\,G_c^{(n)}
\label{2.20}
\end{equation}

\bigskip\noindent 
where $\omega_{ab}$ and $X^{ab}$ are the inverses of $\omega^{ab}$
and $X_{ab}$, and

\begin{equation}
G_a^{(n)}=\sum_{m=0}^n\bigl\{T_a^{(n-m)},\,A^{(m)}\bigr\}_{(q,p)}
+\sum_{m=0}^{n-2}\bigl\{T_a^{(n-m)},\,A^{(m+2)}\bigr\}_{(\eta)}
+\bigl\{T_a^{(n+1)},\,A^{(1)}\bigr\}_{(\eta)}
\label{2.21}
\end{equation}

\bigskip
The general prescription of the usual BFFT method to obtain the
Hamiltonian is to direct use the relations (\ref{2.15}) and
(\ref{2.20}). This works well for system with linear constraints. For
nonlinear theories, where it may be necessary to consider all order
of the iterative process, this calculation might be quite
complicated.  There is an alternative procedure that drastically
simplifies the algebraic work. The basic idea is to
obtain the involutive forms for the initial fields $q$ and $p$
\cite{Banerjee5}. This can be directly achieved from the previous
analysis to obtain $\tilde A$. Denoting these by $\tilde q$ and
$\tilde p$ we have

\begin{equation}
H(q,p)\longrightarrow H(\tilde q,\tilde p)
=\tilde H(\tilde q,\tilde p)
\label{2.22}
\end{equation}

\bigskip\noindent
It is obvious that the initial boundary condition in the BFFT
process, namely, the reduction of the involutive function to the
original function when the new fields are set to zero, remains
preserved. Incidentally we mention that in the cases with linear
constraints, the new variables $\tilde q$ and $\tilde p$ are just
shifted coordinates in the auxiliary coordinate $\eta$ \cite{Ricardo}.

\bigskip
Let us now finally consider the case where the first-class
constraints form an non-Abelian algebra, i.e. 

\begin{equation}
\bigl\{\tilde T_a,\,\tilde T_b\bigr\}=C_{ab}^c\,\tilde T_c
\label{2.23}
\end{equation}

\bigskip\noindent
The quantities $C_{ab}^c$ are the structure constant of the
non-Abelian algebra. These constraints are considered to satisfy the
same previous conditions given by (\ref{2.3}), (\ref{2.4}),
(\ref{2.6}), and (\ref{2.7}). But now, instead of Eqs.
(\ref{2.8})-(\ref{2.10}), we obtain 

\begin{eqnarray}
C_{ab}^c\,T_c&=&\bigl\{T_a,T_b\bigr\}
+\bigl\{T_a^{(1)},T_b^{(1)}\bigr\}_{(\eta)}
\label{2.24}\\
C_{ab}^c\,T_c^{(1)}&=&\bigl\{T_a,T_b^{(1)}\bigr\}
+\bigl\{T_a^{(1)},T_b\bigr\}
\nonumber\\
&&+\,\bigl\{T_a^{(1)},T_b^{(2)}\bigr\}_{(\eta)}
+\bigl\{T_a^{(2)},T_b^{(1)}\bigr\}_{(\eta)}
\label{2.25}\\
C_{ab}^c\,T_c^{(2)}&=&\bigl\{T_a,T_b^{(2)}\bigr\}
+\bigl\{T_a^{(1)},T_b^{(1)}\bigr\}_{(q,p)}
\nonumber\\
&&+\bigl\{T_a^{(2)},T_b^{(0)}\bigr\}_{(q,p)}
+\bigl\{T_a^{(1)},T_b^{(3)}\bigr\}_{(\eta)}
\nonumber\\
&&+\bigl\{T_a^{(2)},T_b^{(2)}\bigr\}_{(\eta)}
+\bigl\{T_a^{(3)},T_b^{(1)}\bigr\}_{(\eta)}
\label{2.26}\\
&&\vdots
\nonumber
\end{eqnarray}

\bigskip\noindent 
The use of these equations is the same as before, i.e., they shall
work iteratively. Equation (\ref{2.24}) gives $T^{(1)}$.  With this
result and Eq. (\ref{2.25}) one calculates $T^{(2)}$, and so on. To
calculate the first correction, we assume it is given by the same
general expression (\ref{2.11}). Introducing it into (\ref{2.24}), we
now get

\begin{equation}
C_{ab}^c\,T_c=\Delta_{ab}+X_{ac}\,\omega^{cd}\,X_{bd}
\label{2.27}
\end{equation}

\bigskip\noindent 
Of course, the same difficulties pointed out with respect the
solutions of Eq.  (\ref{2.12}) also apply here, with the additional
problem of choosing the appropriate structure constants $C_{ab}^c$.

\bigskip
To obtain the embedding Hamiltonian $\tilde H(q,p,\eta)$ one cannot
used the simplified version discussed for the Abelian case, embodied
into Eq. (\ref{2.22}), because the algebra is not strong involutive
anymore. We here start from the fact that the new Hamiltonian $\tilde
H$ and the new constraints $\tilde T$ satisfy the relation

\begin{equation}
\bigl\{\tilde T_a,\,\tilde H\bigr\}=B_a^b\,\tilde T_b
\label{2.28}
\end{equation}

\bigskip\noindent
where the coefficients $B_a^b$ are the same coefficients that may appear
in the consistency condition of the initial constraints, i.e. 

\begin{equation}
\bigl\{T_a,\, H\bigr\}=B_a^b\,T_b
\label{2.29}
\end{equation}

\bigskip\noindent
because in the limit of $\eta\rightarrow0$ both relations
(\ref{2.28}) and (\ref{2.29})coincide. The involutive Hamiltonian
is considered to satisfy the same conditions
(\ref{2.14})-(\ref{2.16}). We then obtain that the general correction
$H^{(n)}$ is given by a relation similar to (\ref{2.20}), but now the
quantities $G_a^{(n)}$ are given by

\begin{eqnarray}
G_a^{(n)}&=&\sum_{m=0}^n\bigl\{T_a^{(n-m)},\,H^{(m)}\bigr\}_{(q,p)}
+\sum_{m=0}^{n-2}\bigl\{T_a^{(n-m)},\,A^{(m+2)}\bigr\}_{(\eta)}
\nonumber\\
&&+\,\,\bigl\{T_a^{(n+1)},\,A^{(1)}\bigr\}_{(\eta)}
-B_a^b\,T_c^{(n)}
\label{2.30}
\end{eqnarray}

\vspace{1cm}
\section{Using the extended BFFT formalism for the O(N) non\-linear
sigma model} 
\renewcommand{\theequation}{3.\arabic{equation}}
\setcounter{equation}{0}

\bigskip
The $O(N)$ nonlinear sigma model is described by the Lagrangian

\begin{equation}
{\cal L}=\frac{1}{2}\,\partial_\mu\phi^A\partial^\mu\phi^A
+\frac{1}{2}\,\lambda\,\bigl(\phi^A\phi^A-1\bigr)
\label{3.1}
\end{equation}

\bigskip\noindent
where the $\mu=0,1$ and $A$ is an index related to the $O(N)$
symmetry group. To obtain the constraints, we calculate the canonical
momenta

\begin{eqnarray}
&&\pi^A=\frac{\partial{\cal L}}{\partial\dot\phi^A}=\dot\phi^A
\label{3.2}\\
&&p=\frac{\partial{\cal L}}{\partial\dot\lambda}=0
\label{3.3}
\end{eqnarray}

\bigskip\noindent
We notice that Eq. (\ref{3.3}) is a primary constraint. In order to
look for secondary constraints we construct the Hamiltonian density

\begin{eqnarray}
{\cal H}&=&\pi^A\dot\phi^A+p\dot\lambda-{\cal L}+\xi p\,,
\nonumber\\
&=&\frac{1}{2}\,\pi^A\pi^A
+\frac{1}{2}\phi^{A\prime}\phi^{A\prime}
-\frac{1}{2}\,\lambda\,\bigl(\phi^A\phi^A-1\bigr)
+\tilde\xi p
\label{3.4}
\end{eqnarray}

\bigskip\noindent
where prime will always mean derivative with respect the space
coordinate $x$.  The velocity $\dot\lambda$ was absorbed in the
Lagrange multiplier $\tilde\xi$ by the redefinition
$\tilde\xi=\xi+\dot\lambda$. The consistency condition for the
constraint $p=0$ leads to another constraint

\begin{equation}
\phi^A\phi^A-1=0
\label{3.5}
\end{equation}

\bigskip\noindent
At this stage, we have two options. The first one, that we shall
consider here, is to introduce the constraint above into the
Hamiltonian by means of another Lagrange multiplier. The result is

\begin{eqnarray}
{\cal H}&=&\frac{1}{2}\,\pi^A\pi^A
+\frac{1}{2}\,\phi^{A\prime}\phi^{A\prime}
-\frac{1}{2}\,\lambda\,\bigl(\phi^A\phi^A-1\bigr)
+\tilde\xi\,p
+\zeta\,\bigl(\phi^A\phi^A-1\bigr)
\nonumber\\
&=&\frac{1}{2}\,\pi^A\pi^A
+\frac{1}{2}\,\phi^{A\prime}\phi^{A\prime}
+\tilde\xi\,p
+\tilde\zeta\,\bigl(\phi^A\phi^A-1\bigr)
\label{3.6}
\end{eqnarray}

\bigskip\noindent
The field $\lambda$ was also absorbed by the Lagrange multiplier
$\zeta$ ($\tilde\zeta=\zeta-\frac{1}{2}\lambda$). The consistency
condition for the constraint (\ref{3.5}) leads to another more
constraint 

\begin{equation}
\phi^A\pi^A=0
\label{3.7}
\end{equation}

\bigskip\noindent
We mention that the consistency condition for this constraint will
give us the Lagrange multiplier $\tilde\zeta$ and no more
constraints.  Since we have absorbed the field $\lambda$, its
momentum $p$ does not play any role in the theory and we can
disregard it by using the constraint relation (\ref{3.3}) in a strong
way. So the constraints in this case are

\begin{eqnarray}
T_1&=&\phi^A\phi^A-1
\nonumber\\
T_2&=&\pi^A\phi^A
\label{3.8}
\end{eqnarray}

\bigskip
The other option we had mentioned was to keep the Lagrange multiplier
$\lambda$ in the theory. To do this, we consider that the constraint
(\ref{3.5}) is already in the Hamiltonian due to the presence of the
term $-\,\frac{\lambda}{2}\,(\phi^A\phi^A-1)$. So, instead of the
Hamiltonian (\ref{3.6}) we use the previous one given by (\ref{3.4})
in order to verify the consistency condition  of the constraint
(\ref{3.5}). It is easily seen that the constraint (\ref{3.7}) is
obtained again, and the consistency condition for it leads to new
constraints, involving $\lambda$. However, the use of the BFFT method
for this set of constraints is not feasible of being applied as was
pointed out in the paper of ref. \cite{Barc1}.

\medskip
Let us extend the phase space by introducing two new variables
($\eta^1,\eta^2$) and consider that $\eta^2$ is the canonical
momentum conjugate to $\eta^1$.

\medskip
Before using the formalism with non-Abelian algebra, it is
instructive to consider the Abelian case first, in order to make some
comparisons. From Eq. (\ref{2.12}), we have

\begin{equation}
X_{12}(x,z)\,X_{21}(y,z)-X_{11}(x,z)\,X_{22}(y,z)
=2\,\phi^A(x)\phi^A(x)\,\delta(x-y)
\label{3.9}
\end{equation}

\bigskip\noindent
where it is understood integrations over the intermediate variable $z$.

\medskip

As one observes, this is just one equation with four quantities to be
fixed. Banerjee {\it et al}. \cite{Banerjee3} have shown that
the choice

\begin{eqnarray}
&&X_{11}(x,y)=2\,\delta(x-y)
\nonumber\\
&&X_{22}(x,y)=-\,\phi^A\phi^A\,\delta(x-y)
\nonumber\\
&&X_{12}(x,y)=X_{21}(x,y)=0
\label{3.10}
\end{eqnarray}

\bigskip\noindent
leads to the following set of linear first-class constraints 

\begin{eqnarray}
\tilde T_1&=&\phi^A\phi^A-1+2\,\eta^1
\nonumber\\
\tilde T_2&=&\phi^A\pi^A-\phi^A\phi^A\,\eta^2
\label{3.11}
\end{eqnarray}

\bigskip\noindent
Another choice that also leads to linear constraints is

\begin{eqnarray}
&&X_{12}(x,y)=2\,\delta(x-y)
\nonumber\\
&&X_{21}(x,y)=\phi^A\phi^A\,\delta(x-y)
\nonumber\\
&&X_{11}(x,y)=X_{22}(x,y)=0
\label{3.12}
\end{eqnarray}

\bigskip\noindent
But this choice is nothing other than the interchange of $\eta^1$ and
$\eta^2$ in Eqs. (\ref{3.11}). 

\medskip
However, if instead of these choices we had considered 

\begin{eqnarray}
&&X_{11}(x,y)=\phi^A\phi^A\,\delta(x-y)
\nonumber\\
&&X_{22}(x,y)=-\,2\,\delta(x-y)
\nonumber\\
&&X_{12}(x,y)=X_{21}(x,y)=0
\label{3.13}
\end{eqnarray}

\bigskip\noindent
it is necessary to go to the second step of the iterative process,
but we mention that is possible to stop the process there
\cite{Barc1}, and the first-class constraints we obtain are

\begin{eqnarray}
\tilde T_1&=&\phi^A\phi^A-1+\phi^A\phi^A\,\eta^1+\eta^2\eta^2
\nonumber\\
\tilde T_2&=&\phi^A\pi^A-2\,\eta^2-2\,\eta^1\eta^2
\label{3.14}
\end{eqnarray}

\bigskip
Let us now consider the BFFT method with a non-Abelian algebra. From
Eq. (\ref{2.27}), we have 

\begin{eqnarray}
X_{12}(x,z)\,X_{21}(y,z)-X_{11}(x,z)\,X_{22}(y,z)
&=&2\,\phi^A(x)\phi^A(x)\,\delta(x-y)
\nonumber\\
&&+\,C_{12}^a(x,y,z)\,T_a(z)
\label{3.15}
\end{eqnarray}

\bigskip\noindent
After some attempts, we find that a convenient choice (as it will
become patent later) for these coefficients is

\begin{eqnarray}
&&X_{11}(x,y)=-\,\delta(x-y)
\nonumber\\
&&X_{22}(x,y)=2\,\delta(x-y)
\nonumber\\
&&X_{12}(x,y)=0=X_{21}(x,y)
\nonumber\\
&&C_{12}^1(x,y,z)=2\,\delta(x-z)\delta(x-y)
\nonumber\\
&&C_{12}^2(x,y,z)=0
\label{3.16}
\end{eqnarray}

\bigskip\noindent
The use of expression (\ref{2.11}), permit us directly obtain the first
corrections for the constraints

\begin{eqnarray}
T_1^{(1)}&=&-\,\eta^1
\nonumber\\
T_2^{(1)}&=&2\,\eta^2
\label{3.17}
\end{eqnarray}

\bigskip\noindent
To calculate the second correction, we have to use the Eq.
(\ref{2.25}). Considering the values we inferred for the structure
constants, we obtain

\begin{equation}
2\,T_1^{(1)}=\bigl\{T_1^{(1)},\,T_2^{(2)}\bigr\}_{(\eta)}
+\bigl\{T_1^{(2)},\,T_2^{(1)}\bigr\}_{(\eta)}
\label{3.18}
\end{equation}

\bigskip\noindent
The combination of Eqs. (\ref{3.17}) and (\ref{3.18}) will lead to an
equation involving $T_1^{(2)}$ and $T_2^{(2)}$. We conveniently take

\begin{equation}
T_2^{(2)}=0
\label{3.19}
\end{equation}

\bigskip\noindent
and directly calculate $T_1^{(2)}$. The result is

\begin{equation}
T_1^{(2)}=-\,\frac{1}{2}\,\bigl(\eta^1\bigr)^2
\label{3.20}
\end{equation}

\bigskip\noindent
For the next corrections we use the Eq. (\ref{2.26}) and the previous
results. We also fix $T_2^{(3)}$ as zero and obtain
$T_1^{(3)}=-\,(\eta^1)^3/6$. The general result can be directly
inferred 

\begin{eqnarray}
T_1^{(n)}&=&-\,\frac{1}{n!}\,\Bigl(\eta^1\Bigr)^n
\nonumber\\
T_2^{(1)}&=&2\,\eta^2
\nonumber\\
T_2^{(n)}&=&0\hspace{.5cm}{\rm for}\hspace{.5cm}n\geq2
\label{3.21}
\end{eqnarray}

\bigskip\noindent
Including all the corrections, we write down the first-class
constraints $\tilde T_a$

\begin{eqnarray}
&&\tilde T_1=\phi^A\phi^A-1
-\sum_{n=1}^\infty\frac{1}{n!}\,\Bigl(\eta^1\Bigr)^n
=\phi^A\phi^A-{\rm e}^{\eta^1}
\label{3.22}\\
&&\tilde T_2=\phi^A\pi^A+2\eta^2
\label{3.23}
\end{eqnarray}

\bigskip\noindent
We observe that the above choices permitted us to sum up the infinite
corrections of $\tilde T_1$ in the exponential term e$^{\eta^1}$. The
first-class constraint algebra is

\begin{eqnarray}
&&\bigl\{\tilde T_1(x),\,\tilde T_1(y)\bigr\}=0
\nonumber\\
&&\bigl\{\tilde T_1(x),\,\tilde T_2(y)\bigr\}
=2\,\tilde T_1(x)\,\delta(x-y)
\nonumber\\
&&\bigl\{\tilde T_2(x),\,\tilde T_2(y)\bigr\}=0
\label{3.24}
\end{eqnarray}

\bigskip
Our next step is the obtainment of the Lagrangian that leads to this
non-Abelian algebra. The canonical Hamiltonian density reads 

\begin{equation}
{\cal H}_c=\frac{1}{2}\,\pi^A\pi^A
+\frac{1}{2}\,\phi^{A\prime}\phi^{A\prime}
-\frac{1}{2}\,\lambda\,\Bigl(\phi^A\phi^A-1\Bigr)
\label{3.25}
\end{equation}

\bigskip\noindent
The corrections for the canonical Hamiltonian are given by Eqs.
(\ref{2.20}) and (\ref{2.30}). Considering that the initial
Hamiltonian $H_c$ and the constraint $T_1$ satisfy
$\{T_1,H_c\}=2\,T_2$ and that $\{T_2,H_T\}$ ($H_T$ is the total
Hamiltonian) will just lead to the evaluation of Lagrange
multipliers, we take $B_1^2=2$ and the remaining coefficients as
zero. The first correction then gives

\begin{equation}
\bigl\{\eta^1(x),\,H^{(1)}_c\bigr\}_{(\eta)}=0
\label{3.26}
\end{equation}

\bigskip\noindent
This permit us to conclude that the general form for the first
correction $H_c^{(1)}$ must be

\begin{equation}
H_c^{(1)}=\int dx\,\alpha_1\,\eta^1
\label{3.27}
\end{equation}

\bigskip\noindent
where the quantity $\alpha_1$ remains to be conveniently fixed.
The second step of the iterative method leads to the equation 

\begin{equation}
4\,\eta^2(x)=-\,\bigl\{\eta^1(x),\,H^{(2)}_c\bigr\}_{(\eta)}
\label{3.28}
\end{equation}

\bigskip\noindent
which permit us to conclude that $H_c^{(2)}$ might have the general
form 

\begin{equation}
H_c^{(2)}=\int dx\bigl(-2\,\eta^2\eta^2
+\alpha_2\,\eta^1\eta^1\bigr)
\label{3.29}
\end{equation}

\bigskip\noindent
where $\alpha_2$ also remains to be fixed. Proceeding successively in this
way, we infer that the general form the correction $H_c^{(n)}$ should
be

\begin{equation}
H_c^{(n)}=\int dx\,\Bigl[\alpha_n\,(\eta^1)^n
-\,\theta(n-2)\,\frac{2\,(-1)^{n-2}}{(n-2)!}\,
(\eta^1)^{n-2}\,(\eta^2)^2\Bigr]
\label{3.30}
\end{equation}

\bigskip\noindent
where the $\theta$-term in the expression above means

\begin{equation}
\theta(n-2)=\left\{
\begin{array}{ll}
1&n>2\\
0&n\leq2
\end{array}
\right.
\label{3.30a}
\end{equation}

\bigskip\noindent
The final form of the involutive Hamiltonian is

\begin{eqnarray}
\tilde H_c&=&H+H_c^{(1)}+H_c^{(2)}+\cdots
\nonumber\\
&=&\int dx\,\Bigl[\frac{1}{2}\,\pi^A\pi^A
+\frac{1}{2}\,\phi^{A\prime}\phi^{A\prime}
-2\,{\rm e}^{-\eta^1}\,\eta^2\eta^2
-\frac{1}{8}\,{\rm e}^{\eta^1}
\eta^{1\prime}\eta^{1\prime}
\nonumber\\
&&\phantom{\int dx\,\Bigl[\frac{1}{2}\,\pi^A\pi^A
+\frac{1}{2}\,\phi^{A\prime}\phi^{A\prime}}
-\frac{1}{2}\,\lambda\,\bigl(\phi^A\phi^A
-{\rm e}^{\eta^1}\bigr)\Bigr]
\label{3.31}
\end{eqnarray}

\bigskip\noindent
where we have taken $\alpha_n=\lambda/2n!$. We have also add the term
$-\frac{1}{8}{\rm e}^{\eta^1}\eta^{1\prime}\eta^{1\prime}$, that does
not spoil the algebra involving $\tilde T_1$ and $\tilde H_c$, in
order to obtain a final covariant Lagrangian. Now, the obtainment of
the Lagrangian is just a matter of direct calculation by means of the
constrained path integral formalism \cite{Faddeev}. We just mention
the result \cite{General}

\begin{equation}
\tilde{\cal L}=\frac{1}{2}\,\partial_\mu\phi^A\partial^\mu\phi^A
+\frac{1}{2}\,\lambda\,\bigl(\phi^A\phi^A-{\rm e}^{\eta^1}\bigr)
-\frac{1}{8}\,{\rm e}^{\eta^1}\partial_\mu\eta^1\partial^\mu\eta^1
\label{3.32}
\end{equation}

\bigskip\noindent
where we notice that there is a kind of a Liouville term in auxiliary
field $\eta^1$.

\vspace{1cm}
\section{The supersymmetric case}
\renewcommand{\theequation}{4.\arabic{equation}}
\setcounter{equation}{0}

\bigskip
The important point to be emphasized in the supersymmetric case is that it is not
possible to obtain a consistent gauge theory by means of an Abelian
algebra \cite{Barc1}. Let us briefly discuss why this occurs. The
constraints for the supersymmetric nonlinear $\sigma$ model (without
considering the Lagrange multipliers) are

\begin{eqnarray}
T_1&=&\phi^A\phi^A-1
\nonumber\\
T_2&=&\phi^A\pi^A
\nonumber\\
T_{3\alpha}&=&\phi^A\psi^A_\alpha
\label{4.1}
\end{eqnarray}

\bigskip\noindent
where the canonical momentum conjugate to $\psi^A_\alpha$ is a
constraint relation 
\footnote{We could have used auxiliary variables to convert these
constraints into first-class too. But this is a trivial operation
that would not lead to any new significant result.} 
that can be eliminated by using the Dirac brackets \cite{Dirac}

\begin{equation}
\bigl\{\psi_\alpha^A(x),\,\psi_\beta^B(y)\bigr\}
=-\,i\,\delta_{\alpha\beta}\,\delta^{AB}\delta(x-y)
\label{4.2}
\end{equation}

\bigskip\noindent
Let us extend the phase-space by introducing the coordinates $\eta_1$,
$\eta_2$, and $\chi_\alpha$. We consider that they satisfy the
following fundamental relations

\begin{eqnarray}
\bigl\{\eta_1(x),\,\eta_2(y)\bigr\}&=&\delta(x-y)
\nonumber\\
\bigl\{\chi_\alpha(x),\,\chi_\beta(y)\bigr\}&=&-\,i
\delta_{\alpha\beta}\,\delta(x-y)
\label{4.3}
\end{eqnarray}

\bigskip
The equation that corresponds to (\ref{2.12}) when fermionic degrees
of freedom are included is

\begin{equation}
\Delta_{ab}+(-1)^{(\epsilon_b+1)\epsilon_d}\,
X_{ac}\,\omega^{cd}\,X_{bd}=0
\label{4.4}
\end{equation}

\bigskip\noindent
where $\epsilon_a=0$ for $a=1,2$ (bosonic constraints) and
$\epsilon_a=1$ otherwise (fermionic ones). Expression (\ref{4.4})
yields the following set of equations

\begin{eqnarray}
&&X_{12}(x,z)\,X_{21}(y,z)-X_{11}(x,z)\,X_{22}(y,z)
\nonumber\\
&&\phantom{X_{12}(x,z)\,}
-iX_{1\,3\alpha}(x,z)\,X_{2\,3\alpha}(y,z)=0
-2\,\phi^A(x)\phi^A(x)\,\delta(x-y)=0
\label{4.5}\\
&&X_{21}(x,z)\,X_{3\alpha\,2}(y,z)-
X_{22}(x,z)\,X_{3\alpha\,1}(y,z)
\nonumber\\
&&\phantom{X_{12}(x,z)\,}
-iX_{2\,3\beta}(x,z)\,X_{3\alpha\,3\beta}(z,y)
-\,\phi^A(x)\psi^A_\alpha(x)\,\delta(x-y)=0
\label{4.6}\\
&&X_{3\alpha\,1}(x,z)\,X_{3\beta\,2}(y,z)
-X_{3\alpha\,2}(x,z)\,X_{3\beta\,1}(y,z)
\nonumber\\
&&\phantom{X_{12}(x,z)\,}
-iX_{3\alpha\,3\gamma}(x,z)\,X_{3\beta\,3\gamma}(z,y)
-\,i\,\delta_{\alpha\beta}\,\phi^A(x)\phi^A(x)\,\delta(x-y)
\label{4.7}\\
&&X_{11}(x,z)\,X_{3\alpha\,2}(y,z)
-X_{12}(x,z)\,X_{3\alpha\,1}(y,z)
\nonumber\\
&&\phantom{X_{12}(x,z)\,}
-iX_{1\,3\beta}(x,z)\,X_{3\alpha\,3\beta}(y,z)=0
\label{4.8}
\end{eqnarray}

\bigskip\noindent
From equation (\ref{4.7}) we are forced to conclude that

\begin{equation}
X_{3\alpha\,3\beta}=i\,\delta_{\alpha\beta}
\,\sqrt{\phi^A\phi^A}\,\delta(x-y)\,,
\label{4.9}
\end{equation}

\bigskip\noindent
and a careful analysis of the remaining equations permit us to
infer that a solution that makes possible to have a an Abelian
first-class constraint algebra with the first step of the method is 

\begin{eqnarray}
X_{11}(x,y)&=&2\,\delta(x-y)
\label{4.10}\\
X_{22}(x,y)&=&-\,\phi^A\phi^A\,\delta(x-y)
\label{4.11}\\
X_{2\,3\alpha}(x,y)&=&\frac{\phi^A\psi^A_\alpha}
{\sqrt{\phi^B\phi^B}}\,\delta(x-y)
\label{4.12}
\end{eqnarray}

\bigskip\noindent
The remaining coefficients are zero. The choice given by equations
(\ref{4.9})-(\ref{4.12}) leads to the following set of first-class
constraints 

\begin{eqnarray}
\tilde T_1&=&\phi^A\phi^A-1+2\eta^1
\nonumber\\
\tilde T_2&=&\phi^A\pi^A-\phi^A\phi^A\,\eta^2
+\frac{\phi^A\psi^A_\alpha\chi_\alpha}
{\sqrt{\phi^B\phi^B}}
\nonumber\\
\tilde T_{3\alpha}&=&\phi^A\psi^A_\alpha
+i\,\sqrt{\phi^A\phi^A}\,\chi_\alpha
\label{4.13}
\end{eqnarray}

\bigskip\noindent
We observe that the Abelian treatment for the supersymmetric
nonlinear sigma model just leads to a nonlocal algebra. It is
opportune to emphasize that this is not a problem related with the
attempt of solving the problem in the first stage of the method. If
we had gone beyond the first step we would obtain that the set of
first-class constraints is \cite{Barc1}

\begin{eqnarray}
\tilde T_1&=&\phi^A\phi^A\bigl(1+\eta^1\bigr)-1
\nonumber\\
\tilde T_2&=&\phi^A\pi^A-2\,\eta^2-2\,\eta^1\eta^2
+\frac{\phi^A\psi^A_\alpha\chi_\alpha}
{\sqrt{\phi^B\phi^B}}
\nonumber\\
\tilde T_{3\alpha}&=&\phi^A\psi^A_\alpha
+i\,\sqrt{\phi^A\phi^A}\,\chi_\alpha
\label{4.14}
\end{eqnarray}

\bigskip\noindent
which are also nonlocal.

\medskip
Let us consider the non-Abelian treatment. The calculations follow
the same lines of the bosonic case, discussed in the previous
section. The first step of the method leads to the equations  

\begin{eqnarray}
&&X_{12}(x,z)\,X_{21}(y,z)-X_{11}(x,z)\,X_{22}(y,z)
-iX_{1\,3\alpha}(x,z)\,X_{2\,3\alpha}(y,z)
\nonumber\\
&&\phantom{X_{12}\,}
-C_{12}^2(x,y,z)\,\phi^A(x)\pi^A(x)
-C_{12\alpha}^3(x,y,z)\,\phi^A(x)\psi^A_\alpha(x)
\nonumber\\
&&\phantom{X_{12}\,}
-2\,\phi^A(x)\phi^A(x)\,\delta(x-y)=0
\label{4.15}\\
&&X_{21}(x,z)\,X_{3\alpha\,2}(y,z)-
X_{22}(x,z)\,X_{3\alpha\,1}(y,z)
-iX_{2\,3\beta}(x,z)\,X_{3\alpha\,3\beta}(z,y)
\nonumber\\
&&\phantom{X_{12}\,}
-C_{23\alpha}^1(x,y,z)\,\bigl(\phi^A(z)\phi^A(z)-1\bigr)
-C_{23\alpha}^2(x,y,z)\,\phi^A(x)\pi^A(x)
\nonumber\\
&&\phantom{X_{12}\,}
-C_{23}^3(x,y,z)\,\phi^A(x)\psi^A_\alpha(x)
-\,\phi^A(x)\psi^A_\alpha(x)\,\delta(x-y)=0
\label{4.16}\\
&&X_{3\alpha\,1}(x,z)\,X_{3\beta\,2}(y,z)
-X_{3\alpha\,2}(x,z)\,X_{3\beta\,1}(y,z)
-iX_{3\alpha\,3\gamma}(x,z)\,X_{3\beta\,3\gamma}(z,y)
\nonumber\\
&&\phantom{X_{12}\,}
-\,C_{3\alpha3\beta}^1(x,y,z)\,\bigl(\phi^A(z)\phi^A(z)-1\bigr)
-C_{3\alpha3\beta}^2(x,y,z)\,\phi^A(z)\pi^A(z)
\nonumber\\
&&\phantom{X_{12}\,}
-C_{3\alpha3\beta}^{3\gamma}(x,y,z)\,\phi^A(z)\psi^A_\gamma(z)
-\,i\,\delta_{\alpha\beta}\,\phi^A(x)\phi^A(x)\,\delta(x-y)=0
\label{4.17}\\
&&X_{11}(x,z)\,X_{3\alpha\,2}(y,z)
-X_{12}(x,z)\,X_{3\alpha\,1}(y,z)
-iX_{1\,3\beta}(x,z)\,X_{3\alpha\,3\beta}(y,z)
\nonumber\\
&&\phantom{X_{12}\,}
-C_{13\alpha}^1(x,y,z)\,\bigl(\phi^A(z)\phi^A(z)-1\bigr)
-C_{13\alpha}^2(x,y,z)\,\phi^A(z)\pi^A(z)
\nonumber\\
&&\phantom{X_{12}\,}
-C_{13\alpha}^{3\beta}(x,y,z)\,\phi^A(z)\psi^A_\beta(z)=0
\label{4.18}
\end{eqnarray}

\bigskip\noindent
We choose the following solution

\begin{eqnarray}
C_{12}^1(x,y,z)&=&2\,\delta(x-z)\delta(z-y)
\nonumber\\
C_{23\alpha}^{3\beta}(x,y,z)
&=&-\,\delta_\alpha^\beta\,\delta(x-z)\delta(z-y)
\nonumber\\
C_{3\alpha3\beta}^1(x,y,z)
&=&-\,i\,\delta_{\alpha\beta}\,\delta(x-z)\delta(z-y)
\nonumber\\
X_{11}(x,y)&=&-\,\delta(x-y)
\nonumber\\
X_{22}(x,y)&=&2\,\delta(x-y)
\nonumber\\
X_{3\alpha3\beta}(x,y)&=&i\,\delta_{\alpha\beta}\,\delta(x-y)
\label{4.19}
\end{eqnarray}

\bigskip\noindent
All the other remaining quantities are considered to vanish. The
first correction of the constraints are then given by

\begin{eqnarray}
T_1^{(1)}&=&-\,\eta^1
\nonumber\\
T_2^{(1)}&=&2\,\eta^2
\nonumber\\
T_{3\alpha}^{(1)}&=&i\,\chi_\alpha
\label{4.20}
\end{eqnarray}

\bigskip
To calculate the next corrections, we consider that $T_2^{(n)}=0$ for
$n\geq2$ and proceed as before. Let us just present the final result

\begin{eqnarray}
\tilde T_1&=&\phi^A\phi^A-{\rm e}^{\eta^1}
\nonumber\\
\tilde T_2&=&\phi^A\pi^A+2\,\eta^2
\nonumber\\
\tilde T_{3\alpha}&=&\phi^A\psi^A_\alpha
+i\,{\rm e}^{\frac{1}{2}\eta^1}\,\chi_\alpha
\label{4.21}
\end{eqnarray}

\bigskip\noindent
This leads to the following non-Abelian algebra

\begin{eqnarray}
\bigl\{\tilde T_1(x),\,\tilde T_2(y)\bigr\}
&=&2\,\tilde T_1(x)\,\delta(x-y)
\nonumber\\
\bigl\{\tilde T_2(x),\,\tilde T_{3\alpha}(y)\bigr\}
&=&-2\,\tilde T_{3\alpha}(x)\,\delta(x-y)
\nonumber\\
\bigl\{\tilde T_{3\alpha}(x),\,\tilde T_{3\beta}(y)\bigr\}
&=&-\,i\,\delta_{\alpha\beta}\,\tilde T_1(x)\,\delta(x-y)
\label{4.22}
\end{eqnarray}

\bigskip\noindent
Other brackets are zero. We emphasize that the with the non-Abelian
treatment, there is no problem of locality in gauging the
supersymmetric nonlinear sigma-model.

\bigskip
The correction of the canonical Hamiltonian can be done as in the
pure bosonic case. The result we find is

\begin{eqnarray}
H_c^{(1)}&=&\int dx\,\Bigl(\alpha_1\,\eta^1
+\beta_{1\alpha}\,\chi_\alpha\Bigr)
\nonumber\\
H_c^{(2)}&=&\int dx\,\Bigl(-2\,\eta^2\eta^2
+\alpha_2\,\eta^1\eta^1
+\beta_{2\alpha}\,\chi_\alpha\eta^1\Bigr)
\nonumber\\
H_c^{(3)}&=&\int dx\,\Bigl(2\eta^1\eta^2\eta^2
+\alpha_3\,\eta^1\eta^1\eta^1
+\beta_{3\alpha}\,\chi_\alpha\eta^1\eta^1\Bigr)
\nonumber\\
&\vdots&
\label{4.23}
\end{eqnarray}

\bigskip\noindent
where $\alpha_n$ and $\beta_{n\alpha}$ are quantities to be fixed.
Calculating other terms, we can figure out that the generic form of
$H_c^{(n)}$ reads

\begin{equation}
H_c^{(n)}=\int dx\,\Bigl[(-1)^{n+1}\theta(n-2)\,
\frac{2}{(n-2)!}\,(\eta^1)^{n-2}\eta^2\eta^2
+\alpha_n\,(\eta^1)^n
+\beta_{n\alpha}\,\chi_\alpha(\eta^1)^{n-1}\Bigr]
\label{4.24}
\end{equation}

\bigskip\noindent
where the $\theta$-term was defined in Eq. (\ref{3.30a}). If one
takes 

\begin{eqnarray}
\alpha_n&=&\frac{\lambda}{2n!}
\nonumber\\
\beta_{n\alpha}&=&-\,i\,\Bigl(\frac{1}{2}\Bigr)^{n-1}\,
\frac{\xi_\alpha}{(n-1)!}
\label{4.26}
\end{eqnarray}

\bigskip\noindent
we get

\begin{eqnarray}
\tilde H_c&=&H_c+H_c^{(1)}+H_c^{(2)}+\cdots
\nonumber\\
&=&\int dx\,\Bigl[\frac{1}{2}\,\pi^A\pi^A
-\frac{1}{2}\,\phi^{A\prime}\phi^{A\prime}
-2\,{\rm e}^{-\eta^1}\eta^2\eta^2
-\frac{1}{8}\,{\rm e}^{\eta^1}\,\eta^{1\prime}\eta^{1\prime}
\nonumber\\
&&\phantom{\int dx\,\Bigl[\frac{1}{2}\,\pi^A\pi^A}
i\,\chi\gamma_5\chi^\prime
-\frac{1}{2}\,\lambda\,\bigl(\phi^A\phi^A-{\rm e}^{\eta^1}\bigr)
\nonumber\\
&&\phantom{\int dx\,\Bigl[\frac{1}{2}\,\pi^A\pi^A}
-\,\xi_\alpha\bigl(\phi^A\psi^A_\alpha+i\,\chi_\alpha\,
{\rm e}^{\frac{1}{2}\eta^1}\bigr)\Bigr]
\label{4.27}
\end{eqnarray}

\bigskip\noindent
where $\lambda$ and $\xi_\alpha$ play the role of Lagrange
multipliers. The obtainment of the corresponding Lagrangian can be
done by integrating out the momenta \cite{Faddeev}. The result is

\begin{eqnarray}
\tilde{\cal L}&=&-\,\frac{1}{2}\,\partial_\mu\phi^A\partial^\mu\phi^A
+\frac{i}{2}\,\bar\psi^A\partial\!\!\!\slash\psi^A
+\frac{1}{8}\,{\rm e}^{\eta^1}\,\partial_\mu\eta^1\partial^1\eta^1
-i\,\bar\chi\partial\!\!\!\slash\chi
\nonumber\\
\phantom{-\,\frac{1}{2}\,\partial_\mu\phi^A\partial^\mu\phi^A}
&&+\frac{1}{2}\,\lambda\bigl(\phi^A\phi^A-{\rm e}^{\eta^1}\bigr)
+\xi_\alpha\,\bigl(\phi^A\psi^A_\alpha
+i\,\chi_\alpha\,{\rm e}^{\frac{1}{2}\eta^1}\bigr)
\label{4.28}
\end{eqnarray}

\vspace{1cm}
\section{Conclusion}
We have used an extension of the BFFT formalism presented by Banerjee
et al. in order to gauge the nonlinear sigma model by means of a
non-Abelian algebra. We have considered the supersymmetric and the
usual cases. We have shown that the supersymmetric case is only
consistently transformed in a first-class theory by means of a
non-Abelian algebra. The usual BFFT treatment would lead to a
nonlocal result.

\vspace{1cm}
\noindent {\bf Acknowledgment:} This work is supported in part by
Conselho Nacional de Desenvolvimento Cient\'{\i}fico e Tecnol\'ogico
- CNPq, Financiadora de Estudos e Projetos - FINEP and Funda\c{c}\~ao
Universit\'aria Jos\'e Bonif\'acio - FUJB (Brazilian Research
Agencies).

\end{document}